\documentclass[aps,eqsecnum,twocolumn,prd,showpacs,floatfix]{revtex4}
\usepackage{graphicx}% Include figure files
\def\bold#1{\setbox0=\hbox{$#1$}%
     \kern-.025em\copy0\kern-\wd0
     \kern.05em\%\baselineskip=18ptemptcopy0\kern-\wd0
     \kern-.025em\raise.0433em\box0 }
\def\slash#1{\setbox0=\hbox{$#1$}#1\hskip-\wd0\dimen0=5pt\advance
         to\wd0{\hss\sl/\/\hss}}
\newcommand{\be}{\begin{equation}}
\newcommand{\ee}{\end{equation}}
\newcommand{\bea}{\begin{eqnarray}}
\newcommand{\eea}{\end{eqnarray}}
\newcommand{\nn}{\nonumber}

\newcommand{\spur}[1]{\not\! #1 \,}
\begin{document}
\begin{titlepage}
\addtolength{\jot}{10pt}

 \preprint{\vbox{\hbox{BARI-TH/07-578 \hfill}
                \hbox{October  2007\hfill} }}

\title{\bf  Identifying  $D_{sJ}(2700)$ through its decay modes}

\author{P. Colangelo$^a$, F. De Fazio$^a$, S. Nicotri$^{a,b}$,  M. Rizzi$^b$ \\}

\affiliation{ $^a$ Istituto Nazionale di Fisica Nucleare, Sezione di Bari, Italy\\
$^b$ Dipartimento di Fisica,  Universit\'a  di Bari, Italy}

\begin{abstract}
We study how to  assign  the recently observed  $D_{sJ}(2700)$ meson
to an appropriate level of the  $c \bar s$ spectrum   by  the analysis of
its  decay modes in final states comprising a light pseudoscalar meson. We use an effective lagrangian approach with
heavy quark and chiral symmetries, obtaining that   the measurement
 of the  $D^* K$ decay width  would allow  to distinguish between two possible
assignments.

 \end{abstract}

\vspace*{1cm} \pacs{13.25.Ft,12.39.Fe,12.39.Hg}

\maketitle
\end{titlepage}

\newpage
\section{Introduction}\label{sec:intro}
Recently, Belle Collaboration collected an  important  new piece
of information on charm spectroscopy from  the first  Dalitz plot analysis of the decay  process $B^+
\to {\bar D}^0 D^0 K^+$  \cite{Brodzicka:2007aa}.  The Dalitz plot
shows an accumulation of events at  $M^2(D^0 {\bar D}^0)=
16-18$ GeV$^2$ and $M^2(D^0 K^+) = 7-8$ GeV$^2$.
This could be interpreted as the overlap of two contributions, a
(horizontal)  band in the ${\bar D}^0 D^0$ channel due to a state possibly identified with
$\psi(4160)$,  and a (vertical) band  in the $D^0 K^+$ channel due to a $c{\bar s}$ state  which does not correspond to  any of the previously observed mesons with this
 quark content. A horizontal band at $M^2(D^0 {\bar
D}^0)\simeq 14.2$ GeV$^2$ is also present, due  to the
production of $\psi(3770)$.

Information  from the Dalitz plot was  enriched by the study
of the individual invariant mass distributions. The
$M^2(D^0 K^+)$  distribution   shows the presence of a resonance,
$D_{sJ}(2700)$, with  parameters:
\bea
M &=& 2708\pm 9 ^{+11}_{-10} \,\, {\rm MeV} \nn\\
\Gamma &=& 108 \pm23 ^{+36}_{-31} \,\, {\rm MeV} \,. \label{par} \eea
Moreover, from  the distribution in the
helicity angle $\theta$,  the angle between the
 $D^0$ momentum and the opposite of the kaon momentum in the ${\bar D}^0 D^0$ rest
 frame,  it was possible to assign the  spin-parity $J^P=1^-$   to  $D_{sJ}(2700)$.
A measurement of the  branching fractions
 \bea
 &&{\cal B}(B^+ \to {\bar D}^0 D^0 K^+) = (22.2 \pm 2.2|_{stat}
{^{+2.6}_{-2.4}} |_{sys}) \times 10^{-4} \nonumber \\
&&{\cal B}(B^+ \to D_{sJ}(2700) {\bar D}^0) \times {\cal
B}(D_{sJ}(2700) \to
 D^0 K^+)  = \nonumber \\ && (11.3 \pm 2.2|_{stat}{^{+1.4}_{-2.8}}|_{sys}) \times 10^{-4}\nn \\
 \label{br} \eea
%at 90$\%$  c.l.
was also carried out.

Belle's observation is just the last one of a series of  discoveries  of open and  hidden charm hadrons
 which have greatly enriched the  charm spectroscopy in the past few years
\cite{reviews}. In  case of open charm, which
$D_{sJ}(2700)$ belongs to, before the   B-factories era  the known $c \bar s$  spectrum consisted of
 four states only:  the pseudoscalar  $D_s(1968)$ and  vector $D_s^*(2112)$ mesons,
corresponding to the  $s$-wave states of the  quark model, and the axial-vector $D_{s1}(2536)$ and tensor
$D_{s2}(2573)$ mesons,   $p$-wave states.
In 2003,  two narrow  resonances: $D_{sJ}(2317)$
 and $D_{sJ}^*(2460)$  were discovered by BaBar \cite{Aubert:2003fg}
 and CLEO \cite{Besson:2003cp} Collaborations,
 respectively,  and later on confirmed by other experiments \cite{otherexp}.
 They were assigned spin-parities $J^P=0^+$ and $J^P=1^+$. Their
identification as  proper $c \bar s$ states has  been  controversial since then
\cite{Colangelo:2004vu};  however, they have the  quantum numbers of the
two states needed to complete the  $p$-wave multiplet, and their radiative decays
occur accordingly, so that  their  interpretation  as ordinary $c{\bar s}$ configurations is  natural \cite{fdf1}.

Last year,  BaBar Collaboration announced the observation  of
another $c{\bar s}$ meson,  $D_{sJ}(2860)$  decaying to $D^0
K^+$ and $D^+ K_S$, with mass $M(D_{sJ}(2860))=2856.6 \pm 1.5 \pm
5.0$ MeV and width $\Gamma(D_{sJ}(2860))=47 \pm 7 \pm 10$ MeV
\cite{Aubert:2006mh}. The quantum numbers of this meson have  still to be assigned, and
  two proposals have been put forward.  One is that this state  is  the first radial
excitation of $D_{sJ}(2317)$, hence with $J^P=0^+$   \cite{van
Beveren:2006st,swanson},  the other one is that the meson has  $J^P=3^-$, an interpretation proposed  in  \cite{Colangelo:2006rq} and supported by  a  lattice QCD
  study \cite{Collaboration:2007nr}.  In  the same  analysis of the $D K$  mass distribution,  BaBar   noticed  a broad structure with  $M=2688 \pm 4 \pm 3$ MeV and  $\Gamma=112
\pm 7 \pm 36$ MeV \cite{Aubert:2006mh},   likely the same  resonance $D_{sJ}(2700)$ found by  Belle.

In this   scenario,  $D_{sJ}(2700)$ needs to be  properly placed in  the spectrum of  mesons with charm and strangeness.
In our study  we investigate  its decay modes
with the aim of finding signatures allowing  to assign the meson to a particular  $c \bar s $  level.
In this respect we follow a different strategy from the one adopted in the framework of quark models   \cite{swanson}, where the assignement is made by comparing the observed mass to the predicted value in a
suitably chosen interquark potential, a procedure which can find difficulties in the cases
where neglected effects, such as the threshold effects, are relevant.

In  Section \ref{frame}  we present a theoretical framework  based upon an effective Lagrangian
describing strong decays of heavy mesons to final states comprising a light pseudoscalar meson,  and
displaying heavy quark and  chiral symmetries. 
We
analyze  the various decay modes and compare the predictions that follow from different assignments.
The implications for the  still unobserved states, including $c \bar q$ mesons, are discussed in
Sections \ref{part1}-\ref{part3}, then our conclusions are presented.

\section{Heavy mesons decays to light pseudoscalar mesons}\label{frame}

The study of properties and decays of hadrons containing a single
heavy quark $Q=c,b$ is suitably carried out  in the  $m_Q \to \infty$
limit, which is formulated in the Heavy Quark Effective Theory
\cite{Neubert:1993mb}. In  such a
limit,  the heavy quark acts as a static colour source for the rest of the hadron;
its spin ${\vec s}_Q$ is decoupled from the total angular momentum of the light degrees of
freedom ${\vec s}_\ell$,  and they are separately conserved. Hadrons
can be classified according to the values of ${\vec s}_\ell$ and
of the total spin   ${\vec J}={\vec s}_Q+{\vec s}_\ell$. In particular,
heavy mesons can be organized in doublets, each one
corresponding to a particular value of $s_\ell$ and parity;  the
members of each doublet differ for the orientation of $s_Q$ with
respect to $s_\ell$ and, in the heavy quark limit, are
degenerate. Mass  degeneracy is broken at order  $1/m_Q$.

For $Q \bar q$ states one can write $\vec s_\ell=\vec s_q+ \vec
\ell$, where $s_q$ is the light antiquark spin and $\ell$ is the
orbital angular momentum of the light degrees of freedom relative
to the heavy quark. The lowest lying $Q \bar q$ mesons correspond
to $\ell=0$ ($s$-wave states of the quark model) with $
s_\ell^P={1 \over 2}^-$. This doublet comprises two states with
spin-parity $J^P=(0^-,1^-)$. For $\ell=1$ ($p$-wave states of the
quark model), it could be either $ s_\ell^P={1 \over 2}^+$ or $
s_\ell^P={3 \over 2}^+$,   the  two corresponding doublets having
$J^P=(0^+,1^+)$ and $J^P=(1^+,2^+)$. The mesons
 with $\ell=2$ ($d$-wave states) are collected either in the
$ s_\ell^P={3 \over 2}^-$ doublet, consisting of  states
with $J^P=(1^-,2^-)$, or in the $ s_\ell^P={5 \over 2}^-$ doublet with  $J^P=(2^-,3^-)$ states. And so on.

 The two  states $D_s(1968)$ and $D_s^*(2112)$ can be identified
with the members of the lowest lying $s_\ell^P={1 \over 2}^-$ doublet.
 The resonances  $D_{s1}(2536)$ and
$D_{s2}(2573)$, together with $D_{sJ}(2317)$
 and $D_{sJ}^*(2460)$,  fill the four $p$-wave levels: in particular, $D_{s2}(2573)$ corresponds to
 $ s_\ell^P={3 \over 2}^+$,  $J^P=2^+$ state, while $D_{sJ}(2317)$ to  $ s_\ell^P={1 \over 2}^+$,  $J^P=0^+$. The two axial-vector mesons $D_{s1}(2536)$ and $D_{sJ}^*(2460)$ could be assigned to a linear
 combination  of   $ s_\ell^P={3 \over 2}^+$ and  $ s_\ell^P={1\over 2}^+$ states, which is allowed
 at $O(1/m_Q)$; however, in case of non-strange charm mesons such a mixing has been measured
 and found to be small \cite{Abe:2003zm,ferrandes}, so that we can identify $D_{s1}(2536)$ and $D_{sJ}^*(2460)$
 with the $J^P=1^+$  $ s_\ell^P={3 \over 2}^+$ and  $ s_\ell^P={1\over 2}^+$ states, respectively.

In the interpretation in \cite{Colangelo:2006rq},    $D_{sJ}(2860)$  corresponds to the $J^P=3^-$ component of the $ s_\ell^P={5 \over 2}^-$ doublet, so that the still unassigned levels
are the two states belonging to the  $ s_\ell^P={3 \over 2}^-$ doublet,  with $J^P=(1^-,2^-)$,
and the partner of  $D_{sJ}(2860)$ in the $ s_\ell^P={5 \over 2}^-$ doublet, which has spin two.

The classification can be continued for higher values of $s_\ell^P$ to describe high spin  mesons; moreover, it can be  replicated for  radial excitations, and it is expected to hold as far as $O(1/m_Q)$
effects are small.

Since the spin-parity of $D_{sJ}(2700)$ has been determined:
$J^P=1^-$, the state fits  either in the doublet with $ s_\ell^P={1
\over 2}^-$ or in the one  with $ s_\ell^P={3 \over 2}^-$.
However,  a  $1^-$ state belonging to the $ s_\ell^P={1
\over 2}^-$ doublet is already known,  $D_s^*(2112)$, so that in this case $D_{sJ}(2700)$ would be a
radial excitation. Combinations of the two cases are not allowed in the heavy quark limit;
a possible role of $1/m_Q$ effects will be discussed below.
Therefore,  two  possibilities  must  be considered:
 \begin{itemize}
 \item $D_{sJ}(2700)$ belongs to the doublet  with $ s_\ell^P={1
\over 2}^-$ and is the first radial excitation; we  denote this
state as $D_s^{* \prime}$;
\item $D_{sJ}(2700)$ is the low lying state with $ s_\ell^P={3 \over 2}^-$,
denoted as $D_{s1}^{* }$.
\end{itemize}

These two possibilities can be conveniently analyzed  adopting a formalism which represents
 the various doublets by effective fields being $4 \times 4$ matrices \cite{traceformalism}.
 The two doublets of  interest,  with  $ s_\ell^P={1 \over 2}^-$ denoted by $H_a$, and  with
$ s_\ell^P={3 \over 2}^-$  denoted by $X_a$ ($a$ is a light flavour index) are  given by:
\bea  H_a &=&
\frac{1+{\rlap{v}/}}{2}[P_{a\mu}^*\gamma^\mu-P_a\gamma_5]
\label{neg} \nonumber \\
X_a^\mu &=&\frac{1+{\rlap{v}/}}{2}
\nonumber \\&\times& \left\{ P^{*\mu\nu}_{2a} \gamma_5 \gamma_\nu
-P^{*\prime}_{1a\nu} \sqrt{3 \over 2}  \left[ g^{\mu
\nu}-{\gamma^\nu \over 3} (\gamma^\mu-v^\mu) \right] \right\}
\nonumber \\ \label{doublets} \eea
with $v$ the meson four-velocity.
Notice that $H_a$ describes  the fundamental   $ s_\ell^P={1 \over 2}^-$ doublet;  the doublet
corresponding to the  first radial excitations is
described by an identical structure $H_a^\prime$. The various
operators in (\ref{doublets}) annihilate mesons of four-velocity $v$ which is
conserved in strong interaction processes: the heavy field
operators  contain a factor $\sqrt{m_P}$ and have dimension $3/2$.

The interaction of these heavy mesons  with the octet of light
pseudoscalar mesons, introduced using the fields
 $\displaystyle \xi=e^{i {\cal M} \over
f_\pi}$, with the matrix ${\cal M}$ containing
$\pi, K$ and $\eta$ fields:
\begin{equation}
{\cal M}= \left(\begin{array}{ccc}
\sqrt{\frac{1}{2}}\pi^0+\sqrt{\frac{1}{6}}\eta & \pi^+ & K^+\nonumber\\
\pi^- & -\sqrt{\frac{1}{2}}\pi^0+\sqrt{\frac{1}{6}}\eta & K^0\\
K^- & {\bar K}^0 &-\sqrt{\frac{2}{3}}\eta
\end{array}\right)
\end{equation}
and $f_{\pi}=132$ MeV,
 can be described by an effective Lagrangian  invariant under chiral transformations
 of the light fields and
 heavy-quark spin-flavour transformations of the heavy fields \cite{hqet_chir}.
At the leading order in the heavy quark mass and light meson
momentum expansion the decays  $F \to H M$ $(F=H^\prime$ and $X$,  and $M$
a light pseudoscalar meson) can be described by the Lagrangian
interaction  terms \cite{hqet_chir}:
\bea &&{\cal L}_{H^\prime} = \,  \tilde g \, Tr [{\bar H}_a^\prime
H_b \gamma_\mu \gamma_5 {\cal
A}_{ba}^\mu ] \label{lh} \\
&&{\cal L}_X =  {k^\prime \over \Lambda_\chi}Tr[{\bar H}_a X^\mu_b
(i D_\mu {\spur {\cal A}}+i{\spur D} { \cal A}_\mu)_{ba} \gamma_5
] + h.c. \hskip 0.7 cm
 \label{lx}  \eea
where
${\cal A}_{\mu ba}=\frac{i}{2}\left(\xi^\dagger\partial_\mu
\xi-\xi
\partial_\mu \xi^\dagger\right)_{ba} $
and $D$ is the covariant derivative
$D_{\mu ba}=-\delta_{ba}\partial_\mu+{\cal V}_{\mu ba}$
with
$V_{\mu b a}=\frac{1}{2}\left(\xi^\dagger\partial_\mu \xi
+\xi\partial_\mu \xi^\dagger\right)_{ba}$.
$\Lambda_\chi$ is  the chiral symmetry-breaking scale
and ${\tilde g}$ , $k^\prime$ are effective couplings. We use $\Lambda_\chi = 1 \, $ GeV, while at present the value of the couplings  is not known  neither
from experiment  nor from theoretical considerations.

The widths of  the
allowed decay  modes of $D_{sJ}(2700)$ for  the two possible interpretations
$D_s^{*\prime}$  or   $D_{s1}^*$ can be computed using (\ref{lh}) and (\ref{lx}). Since a heavy
vector state can decay to a light pseudoscalar and a heavy pseudoscalar or vector
meson, we consider the modes:
$D_{sJ} (2700) \to D^+K_S$,  $D^0 K^+$, $D_s \eta$
and
$D_{sJ} (2700) \to D^{*+} K_S$, $ D^{*0} K^+$, $D_s^* \eta$,
and compute  the ratios of decay widths:
 \bea
R_1 &=& \Gamma(D_{sJ}
\to D^* K) \over \Gamma(D_{sJ} \to D K) \nn \\
R_2 &=& \Gamma(D_{sJ}
\to D_s \eta) \over \Gamma(D_{sJ} \to D K) \label{ratios} \\
R_3 &=& \Gamma(D_{sJ} \to D^*_s \eta) \over \Gamma(D_{sJ} \to D K)
\nn \eea
(with  $D^{(*)}K=D^{(*)+} K_S +D^{(*)0} K^+$)
for both the assignments of  $s_\ell^P$ to $D_{sJ}(2700)$.
 These ratios are useful
 to discriminate between  the two assignments; moreover, the method is sensible, since
in the  ratios (\ref{ratios})  the dependence on the (unknown)
effective couplings drops out, and  the predictions  are
model independent.

Let us first identify $D_{sJ}(2700)$ with $D_s^{*\prime}$.
Using the effective Lagrangian (\ref{lh}), the
$D_s^{*\prime}$ decay widths  can be written as follows:
\bea \Gamma(D_s^{*\prime } \to D_a P_a) &=& C_P {{\tilde g}^2 \over 6
\pi
f_\pi^2}{M_{D_a} \over M_{D_s^{*\prime}}} |\vec q|^3 \nn \\
\Gamma(D_s^{*\prime } \to D_a^* P_a) &=& C_P {{\tilde g}^2 \over 3
\pi f_\pi^2}{M_{D_a^*} \over M_{D_s^{*\prime}}} |\vec q|^3
\label{wid1} \eea  where the light flavour index $a=u,d,s$ identifies
 $D_a^{(*)}=D^{(*)0},D^{(*)+},D_s^{(*)}$, respectively,
and $P_a$ represents a light pseudoscalar meson with quark content
$\bar s  a$ ($P_a=K^+$, $K_{S(L)}$, $\eta$). $C_P$ is a
coefficient depending on the $P$  meson:  $C_{K^+}=1$, $C_{K_S}=\displaystyle{1 \over 2}$ and
$C_\eta=\displaystyle{2 \over 3}$; the modulus of the three momentum $\vec q$
reads:
$|\vec q|= \lambda^{1/2}(M^2_{D_s^{*\prime}},M^2_{D^{(*)}_a},M_{P_a}^2)/2 M_{D_s^{*\prime}}$.

On the other hand, if $D_{sJ}(2700)$ is identified with $D_{s1}^*$,  from  the Lagrangian (\ref{lx}) we obtain:
 \bea
&&\Gamma(D_{s1}^* \to D_a P_a) =\nn \\ && C_P {16   \over 9 \pi
f_\pi^2}\left( {k^\prime \over \Lambda_\chi}\right)^2
{M_{D_a} \over M_{D_{s1}^*}}\left[M^2_{P_a}+|\vec q|^2 \right] |\vec q|^3 \nn \\
&&\Gamma(D_{s1}^* \to D_a^* P_a) =\nn \\  & & C_P {2   \over 9 \pi
f_\pi^2}\left( {k^\prime \over \Lambda_\chi}\right)^2{M_{D_a^*}
\over M_{D_{s1}^*}}\left[M^2_{P_a}+|\vec q|^2 \right] |\vec q|^3
\,.\,\,\,\,\,\,\,\label{wid2} \eea

These two sets of expressions produce different values for the ratios (\ref{ratios}), as one can appreciate
considering  Table \ref{table:ratios}  where the numerical results are collected (with the errors
obtained considering  the uncertainty in the $D_{sJ}(2700)$ mass).

\begin{table}[h]
\caption{Ratios $R_i$ for $D_{sJ}(2700)$ identified as   $D_s^{*\prime}$ or   $D_{s1}^*$.}
\label{table:ratios}
\begin{center}
\begin{tabular}{|c|c | c| c|} \hline
     & $R_1 \times 10^{2} $ & $R_2 \times 10^{2} $ & $R_3 \times 10^{2}$ \\ \hline
  $D_s^{*\prime}$ & $91 \pm 4$ & $20 \pm 1$ & $5 \pm 2$  \\ \hline
  $D_{s1}^*$ & $4.3 \pm 0.2$ & $16.3 \pm  0.9$ & $0.18 \pm 0.07 $ \\
  \hline
\end{tabular}
\end{center}
\end{table}

The ratios $R_1$ and $R_3$ are very different if  $D_{sJ}(2700)$ is  $D_s^{*\prime}$ or   $D_{s1}^*$,
so that  the measurements of these ratios  allow to properly identify  the $D_{sJ}(2700)$.
Since  $R_3$ is  small,
  the  $D^*K$ decay mode is the main  signal that must be investigated in order to
distinguish between the two  possible assignments for $D_{sJ}(2700)$.

It is worth noticing that the results in  Table \ref{table:ratios}  are different from the ones obtained in 
ref.\cite{Zhang:2006yj}  using the $^3P_0$ model,  a quark model with harmonic oscillator meson wave functions, where it turns out that
 the $DK$ mode is suppressed if $D_{sJ}(2700)$ is identified with $D_s^{*\prime}$.
  Moreover, in the  $D_{s1}^*$ identification,  more reduction 
 of the $D^*K$ signal than obtained in   
 ref.\cite{Wei:2006wa} is expected.
 
From the ratios in Table  \ref{table:ratios}, assuming that the width of $D_{sJ}(2700)$ is saturated by decay modes with a heavy
meson and a light pseudoscalar meson in the final state, we can determine the coupling constants
appearing in  (\ref{lh}) and (\ref{lx}), in correspondence of the two possible assignments for $D_{sJ}(2700)$. This assumption is reasonable, since decay modes with more than one light
pseudoscalar meson, or a light vector meson in the final states are severely phase-space suppressed.

Identifying $D_{sJ}(2700)$ with $D_s^{*\prime}$ we obtain: \be
\tilde g= 0.26 \pm 0.05\label{gtilde} \ee while if
 $D_{sJ}(2700)$ is $D_{s1}^*$ we obtain \be k^\prime=0.14 \pm
0.03 \,. \label{kprimo} \ee

These two values are similar to the results obtained for analogous
coupling constants appearing in the effective heavy quark chiral
Lagrangians  \cite{Colangelo:1995ph}. Using the results
(\ref{gtilde})-(\ref{kprimo})  the  branching fractions of the
various decay modes can be computed; the  results are collected  in
Table \ref{table:br-s}. The errors in eqs.(\ref{gtilde}) and
(\ref{kprimo}), as well as those in Table \ref{table:br-s}, are
obtained from the uncertainties in $M_{D_{sJ}}$ and
$\Gamma(D_{sJ})$.

\begin{table*}
\caption{$D_{sJ}(2700)$  branching fractions  in correspondence to
the two  assignments.} \label{table:br-s}
\begin{center}
\begin{tabular}{|c|c | c| c|c|c|c|} \hline
     & ${\cal B}(D_{sJ} \to D^0K^+)$ & ${\cal B}(D_{sJ} \to D^+K_S)$ &
     ${\cal B}(D_{sJ} \to D_s\eta)$ &
     ${\cal B}(D_{sJ} \to D^{*0}K^+)$ & ${\cal B}(D_{sJ} \to D^{*+}K_S)$ &
     ${\cal B}(D_{sJ} \to D_s^*\eta)$
     \\
  \hline
  $D_s^{*\prime}$ & $(24 \pm 14 )\%$ & $(12 \pm 7.0)\%$ & $( 7\pm 4 )\%$
  & $( 22\pm  13)\%$ & $( 10\pm 6 )\%$ & $( 1.7\pm 1.2)\%$\\ \hline
  $D_{s1}^*$ & $( 44 \pm 25 )\%$ & $ (21\pm 12 )\%$ & $(11\pm 6)\%$
   & $( 1.9\pm  1.1)\%$ & $ (0.9\pm 0.5 )\%$ & $(0.12\pm 0.09 )\%$\\
  \hline
\end{tabular}
\end{center}
\end{table*}

A comment is in order about the accuracy of the results. The effective Lagrangians
(\ref{lh})-(\ref{lx}) coincide with the first terms of an expansion in the light pseudoscalar meson momenta. Since in the decays we have considered, such momenta are not very small, one should in principle add other terms, which should be weighted by new, unknown cupling constants. Analogusly, corrections to the heavy quark limit , which is also used in   (\ref{lh})-(\ref{lx}), could be considered, by adding ${\cal O}(1/m_Q)$ terms, which would contain new unknown constants as well \cite{ferrandes}. We cannot assess the role and the size of such corrections on general grounds, however we expect that they would largely cancel out in the ratios of widths. On the other hand, a mixing between the two $J^P=1^-$,
$s_\ell^P={1\over2}^-$, ${3\over2}^-$ states, which is possible at ${\cal O}(1/m_Q)$, would involve a mixing angle that could be fixed by measuring the ratios in Table  \ref{table:ratios} and comparing the experimental results with the ratios computed in the  heavy quark limit.

\section{Partners without strangeness}\label{part1}

It is interesting to  study
the charmed mesons with the same quantum numbers as
$D_{sJ}(2700)$, but with  a different light quark flavour. These
states are a charged charmed meson and a neutral one, denoted as  $D_J^+$ and $D_J^0$, respectively.  They have not
been observed yet,  so that their masses are unknown. We
fix such masses to $2600 \pm 50$ MeV by  the reasonable assumption that
$D_{sJ}(2700)$ is heavier by an amount of the size   of the strange quark mass.

Allowed decay modes for $D_J^+(2600)$ are:
$D_J^+  \to D^0 \pi^+$, $D^+ \pi^0$,  $D_s
{\bar K}_{S(L)}^0$, $D^+ \eta$,  and  $D_J^+  \to D^{*0}\pi^+$, $D^{*+}\pi^0$, $D^{*+} \eta$,
while for $D_{J}^0$ they are:
 $D_J^0  \to D^+ \pi^-$, $D^0 \pi^0$, $D_s  K^-$, $D^0 \eta$ and $D_J^0
\to D^{*0}\pi^0$, $D^{*+}\pi^-$, $D^{*0}\eta$;
 the corresponding widths are
obtained using eq. (\ref{wid1}-\ref{wid2}),  and depend on  the
possible identification of  $D_{J}^{+(0)}$.  
The states  having
$s_\ell=\displaystyle{1 \over 2}^-$ are  denoted as
$D^{*\prime+(0)}$ and are radial excitations, while the states having $s_\ell=\displaystyle{3
\over 2}^-$  are  denoted as $D^{*+(0)}_1$.

Using  the effective coupling
constants ${\tilde g}$ and $k^\prime$  in (\ref{gtilde}),(\ref{kprimo}),
we obtain:

\be \Gamma(D^{*\prime+(0)})=(128 \pm 61) \,\,{\rm MeV} \ee \be
 \Gamma(D_1^{*+(0)})= (85 \pm 46)\,\,{\rm MeV}
 \label{totwid-ud-part}\ee
so that  the $c \bar q$  partners have widths wich are different
in the case of
 the two  assignments, although with a sizeable uncertainty.   Since the mesons are not very broad,
it should be possible to observe them.  The predicted
 branching fractions, collected  in Table  \ref{table:djpiu}, confirm
that the two assignments produce  different results.  In the
identification with the state $D^{*\prime}$, the mode
$D^{*\prime} \to D^* \pi$ has  the largest branching
fraction, while in the second hypothesis, i.e. $D_J=D_1^*$, the
mode with the largest branching ratio is $D^*_1 \to D \pi$.

\begin{table*}
\caption{Branching ratios of  the $c \bar q$ partners of
$D_{sJ}(2700)$,  $D_{J}^+$ and  $D_{J}^0$, for the two possible
assignments; the mass of the states is fixed to $2600 \pm 50$ MeV.
} \label{table:djpiu}
\begin{center}
\begin{tabular}{|c|c | c| c|c|c|c|c|} \hline
     & ${\cal B}(D_{J}^+ \to D^0\pi^+)$ & ${\cal B}(D_{J}^+ \to D^+\pi^0)$ & ${\cal B}(D_J^+
     \to D_s {\bar K}_S^0)$& ${\cal B}(D_{J}^+ \to D^+\eta)$ & ${\cal B}(D_{J}^+ \to D^{*0}\pi^+)$
     & ${\cal B}(D_{J}^+ \to D^{*+}\pi^0)$& ${\cal B}(D_{J}^+ \to D^{*+}\eta)$
     \\
  \hline
  $D^{*\prime +}$ & $(27.0 \pm 2.1 )\%$ & $( 13.3 \pm 1.0)\%$ & $( 2.3\pm 0.8 )\%$
  &$( 5.3\pm 1.0)\%$  & $(32.4 \pm 0.8 )\%$ &$( 16.1\pm 0.4)\%$ & $( 1.2\pm 1.8)\%$\\ \hline
  $D_{1}^{*+}$ & $( 51.1 \pm 3.5  )\%$ & $ (25.0 \pm 1.7 )\%$ & $( 4.0\pm 1.4)\%$
  & $ (11.8\pm  2.1)\%$ & $(2.7 \pm 0.1 )\%$ &$ (1.3\pm  0.1)\%$   & $ (0.1\pm  0.2)\%$\\
  \hline
  \hline
     & ${\cal B}(D_{J}^0 \to D^+\pi^-)$ & ${\cal B}(D_{J}^0 \to D^0\pi^0)$ & ${\cal B}(D_J^0
     \to D_s  K^-)$& ${\cal B}(D_{J}^0 \to D^0\eta)$
    & ${\cal B}(D_{J}^0 \to D^{*+}\pi^-)$ & ${\cal B}(D_{J}^0 \to
    D^{*0}\pi^0)$&
    ${\cal B}(D_{J}^0 \to D^{*0}\eta)$\\
  \hline
  $D^{*\prime 0}$ & $(26.5 \pm 2.1 )\%$ & $( 13.5 \pm 1.1)\%$ & $( 4.9\pm 1.6 )\%$
  & $( 5.5\pm 1.0)\%$ &$(32.0 \pm 0.7 )\%$& $( 16.3\pm 0.4)\%$  & $( 1.3\pm 1.9)\%$\\ \hline
  $D_{1}^{*0}$ & $( 49.7 \pm 3.3  )\%$ & $ (25.6\pm 1.8 )\%$ & $( 8.3\pm 2.8)\%$
   & $(12.3 \pm 2.0 )\%$ &$ (2.6\pm  0.1)\%$& $(1.3 \pm 0.1 )\%$   & $(0.1 \pm 0.2 )\%$\\
  \hline
\end{tabular}
\end{center}
\end{table*}

\section{Spin partners}\label{part2}

Since in the heavy quark limit the heavy mesons are collected  in
doublets with a definite value of $s_\ell^P$,
the state $D_{sJ}(2700)$ has a partner  from which it differs only for the value of the total
spin.

The partner of  $D_s^{*\prime}$
($s_\ell^P=\displaystyle{1 \over 2}^-$) has  $J^P=0^-$;
it is denoted   $D_s^\prime$,   the first radial excitation of $D_s$.
On the other hand, the partner of $D_{s1}^*$
($s_\ell^P=\displaystyle{3 \over 2}^-$) has  $J^P=2^-$: we
 refer to this state as to $D_{s2}^*$.
In both cases,  the  decay modes
$D_s^\prime$, $D_{s2}^*$ $\to D^{*0} K^+$, $ D^{*+} K^0_{S(L)}$, $ D^{*}_s \eta$,
are permitted.
Using the effective Lagrangians (\ref{lh})-(\ref{lx}) we find:
\bea && \Gamma(D_s^{\prime } \to D_q^* P_q) = C_P {{\tilde g}^2
\over 2\pi f_\pi^2}{M_{D_q^*} \over M_{D_s^{\prime}}} |\vec q|^3 \nn
\\
&&\Gamma(D_{s2}^* \to D_q^* P_q) =\nn \\&& C_P {4   \over 6 \pi
f_\pi^2}\left( {k^\prime \over \Lambda_\chi}\right)^2{M_{D_q^*}
\over M_{D_{s2}^*}}\left[M^2_{P_q}+|\vec q|^2 \right] |\vec q|^3 \,\,\, .
\nn \\ \label{widspin-part}\eea

In the heavy quark limit, these partners are degenerate,  hence, in
the numerical analysis we assign them the same mass as
$D_{sJ}(2700)$.  Using the obtained values for $\tilde g$ and $k^\prime$,  we get: \bea
\Gamma(D_s^\prime)&=& (70 \pm 30) \,\,{\rm MeV} \nn \\
\Gamma(D_{s2}^*)&=& (12 \pm 5) \,\,{\rm MeV}
\label{gammatotspinpartners} \eea
and the  branching fractions in Table \ref{table:spin-partner}.
In  the two assignments the spin partners   differ for their  decay width.

\begin{table*}
\caption{Branching ratios of  the spin partner of  $D_{sJ}$ for  the
two quantum number assignments.} \label{table:spin-partner}
\begin{center}
\begin{tabular}{|c|c | c| c|} \hline
  & ${\cal B} (D_s^{\prime}(D_{s2}^*)\to D^{*0}K^+)$ & ${\cal B}(D_s^{\prime}(D_{s2}^*)\to D^{*+}K_S)$ &
     ${\cal B}(D_s^{\prime}(D_{s2}^*)\to D_s^*\eta)$
     \\\hline
  $D_s^{\prime} \, (J^P=0^-)$ & $( 50.0\pm  0.5)\%$ & $( 23.7\pm 0.2 )\%$ & $( 2.6\pm 0.9)\%$  \\ \hline
  $D_{s2}^* \, (J^P=2^-)$ & $( 49.8\pm  0.6)\%$ & $ (23.6\pm 0.2 )\%$ & $(3.1\pm 1.0 )\%$ \\
  \hline
\end{tabular}
\end{center}
\end{table*}

\section{$D_{sJ}(2700)$  decay constant}\label{part3}

We conclude our study observing that, since $D_{sJ}(2700)$ has been discovered  through
the production in  $B^+ \to {\bar D}^0 D_{sJ}(2700)$, it is possible to estimate its
decay constant $f_{D_{sJ}}$  defined as:
 \be
<0|{\bar s}\gamma_\mu(1-\gamma_5)c|D_{sJ}(p,\epsilon)>=f_{D_{sJ}}
\, M_{D_{sJ}} \, \epsilon_\mu \,. \label{fdsj} \ee

The effective Hamiltonian governing $b \to c {\bar c} s$
transitions: \be H_{eff}={G_F \over \sqrt{2}} V_{cb} V_{cs}^*
\left[ C_1(\mu) O_1 + C_2(\mu) O_2 \right] \,\,,\label{heff}
\ee where $C_1$ and $C_2$ are Wilson coefficients and penguin
operators have been neglected, involves the operators:
\bea O_1 &=& [{\bar c} \gamma_\mu
(1-\gamma_5)b] [{\bar s} \gamma^\mu (1-\gamma_5) c] \nn \\
O_2 &=& [{\bar c} \gamma_\mu (1-\gamma_5)c] [{\bar s} \gamma^\mu
(1-\gamma_5) b] \,. \label{operators} \eea
In principle, the dependence
of the Wilson coefficients $C_i$ on the scale $\mu$   should cancel against the $\mu$-dependence of
the matrix elements of the $O_i$. As it is well known, the calculation of
matrix elements such as $\langle D_{sJ}D|H_{eff}|B\rangle$ is a difficult task.  The simplest evaluation  can be obtained by
  naive factorization
\cite{Neubert:1997uc},
in which
the  $B^+ \to {\bar D}^0 D_{sJ}(2700)$ decay amplitude is written  as:
 \bea {\cal A}(B^+
\to {\bar D}^0 D_{sJ}(2700))=a_1 {G_F \over \sqrt{2}} V_{cb}
V_{cs}^*  \times \nn \\<{\bar D}^0(v^\prime)| {\bar b} \gamma^\mu
(1-\gamma_5) c|B(v)> \times \nn \\ <D_{sJ}(p,\epsilon)| {\bar c}
\gamma_\mu(1-\gamma_5)|0> \label{fact-amp} \eea
with  $a_1=C_1+C_2 / 3$, $G_F$ the Fermi constant and $V_{ij}$ elements of the Cabibbo
Kobayashi Maskawa mixing matrix; we use $a_1=1.1$ together with the values quoted
in \cite{PDG} for $V_{ij}$.

In the heavy quark limit, the first matrix element in
(\ref{fact-amp}) can be written in terms of the Isgur-Wise
function $\xi(v \cdot v^\prime)$ \cite{Neubert:1993mb}: \be <{\bar
D}^0(v^\prime)| {\bar b} \gamma^\mu (1-\gamma_5) c|B(v)>=\xi(v
\cdot v^\prime)\sqrt{M_B M_D}(v+v^\prime)^\mu \,\,\,  \label{btod} \ee
 with $v$ and $v^\prime$  the four-velocities of the heavy
mesons.
 The linear parametrization:
 %\footnote{An analytic expression for the  Isgur-Wise function:
 %$\displaystyle \xi(v \cdot v^\prime)=\left({2 \over v \cdot v^\prime +1}\right)^{3/2}$ has been proposed in   F.~Jugeau, A.~Le Yaouanc, L.~Oliver and J.~C.~Raynal,
  %``Explicit Form Of The Isgur-Wise Function In The Bps Limit,''
%  Phys.\ Rev.\  D {\bf 74}, 094012 (2006).
 % %%CITATION = PHRVA,D74,094012;%%.
  %} :
 \be
 \xi(v \cdot v^\prime)=1-{ \rho}^2(v \cdot v^\prime -1) \label{xi}
\ee
involves  the slope  ${ \rho}^2$, for which  various determinations are available:
  ${ \rho}^2=0.83^{+0.15+0.24}_{-0.11-0.01}$  from lattice QCD \cite{slope1},
    ${ \rho}^2=1.179\pm0.048\pm 0.028$  from a recent BaBar measurement \cite{slope2},
   ${ \rho}^2=1.26\pm0.16\pm 0.11$   from a Belle measurement \cite{Abe:2001cs}.
The Heavy Flavour Averaging Group quotes ${ \rho}^2=1.23\pm0.05$ from $B \to D^*$ decays, and
${ \rho}^2=1.17\pm0.18$ from $B \to D$ transitions \cite{hfag}: we use this last value.
 Computing ${\cal B}(B^+
\to {\bar D}^0 D_{sJ}(2700))$ by naive factorization,
considering the branching fraction ${\cal B}(D_{sJ}(2700) \to D^0 K^+)$ in Table
\ref{table:br-s} for the two possible assignments to $D_{sJ}(2700)$,  and comparing
the result  to  (\ref{br}), we obtain:
\bea f_{D_s^{*\prime}} &=& (243 \pm 41) \, {\rm MeV}\nn \\
f_{D_{s1}^*} &=& (181 \pm 30) \, {\rm MeV} \,\label{ff} \eea
depending on the   assignements. In both cases, the leptonic constant turns out to be sizeable, similar
  to  the measured  $D_s$ decay constant: $f_{D_s}=274
\pm 13 \pm 7$ MeV \cite{Pedlar:2007za}.

\section{Conclusions} \label{sec:concl}
We have discussed possible ways  to distinguish between  two assignments for the state $D_{sJ}(2700)$.
Our  result
is that the decay mode to $D^*K$ has very different branching
ratios in the two possible assignments,  so that a measurement of  such a branching fraction
would be useful  to shed light on the  identification of $D_{sJ}(2700)$.

We have also obtained
several predictions, namely   the effective couplings
governing the strong decays of $D_{sJ}$ to $D(D^*)$ and a light
pseudoscalar meson, which are different according to the
 adopted interpretation. We have also  obtained predictions for
non strange partners of $D_{sJ}$, for which we also found that the
decay to $D^* \pi$ has the largest branching ratio, and
predictions for the spin partners.
Further reasearch on   $D_{sJ}(2700)$ according to these suggestions would be useful
to complete our understanding of the open charm meson spectrum.

%\vspace*{1cm}
\acknowledgments
We thank A. Palano and T.N. Pham  for discussions.
This work was supported in part by the EU contract No. MRTN-CT-2006-035482,
"FLAVIAnet".

%*****************
% \clearpage
 

\begin{thebibliography}{99}

\bibitem{Brodzicka:2007aa}
K.~Abe {\it et al.}  [Belle Collaboration],
  %``Observation of a new D/sJ meson in B+ --> anti-D0 D0 K+ decays,''
  arXiv:hep-ex/0608031;
  J.~Brodzicka {\it et al.}  [Belle Collaboration],
  %``Observation of a new D_sJ meson in B+->D0BD0K+ decays,''
  arXiv:0707.3491 [hep-ex].
  %%CITATION = ARXIV:0707.3491;%%

\bibitem{reviews}
For reviews see:
E.~S.~Swanson,
  %``The new heavy mesons: A status report,''
  Phys.\ Rept.\  {\bf 429}, 243 (2006);
  %%CITATION = PRPLC,429,243;%%
   P.~Colangelo, F.~De Fazio, R.~Ferrandes and
S.~Nicotri,
  %``New open and hidden charm spectroscopy,''
  arXiv:hep-ph/0609240
  %%CITATION = HEP-PH/0609240;%%
and
  %``Puzzles in charm spectroscopy,''
  arXiv:0706.0053 [hep-ph].
  %%CITATION = ARXIV:0706.0053;%%


\bibitem{Aubert:2003fg}
  B.~Aubert {\it et al.}  [BABAR Collaboration],
  %``Observation of a narrow meson decaying to $D_s^+ \pi^0$ at a mass of
  %2.32-GeV/c$^2$,''
  Phys.\ Rev.\ Lett.\  {\bf 90}, 242001 (2003).
  %%CITATION = PRLTA,90,242001;%%

\bibitem{Besson:2003cp}
D.~Besson {\it et al.}  [CLEO Collaboration],
  %``Observation of a narrow resonance of mass 2.46-GeV/c**2 decaying to  D/s*+
  %pi0 and confirmation of the D/sJ*(2317) state,''
  Phys.\ Rev.\  D {\bf 68}, 032002 (2003)
  [Erratum-ibid.\  D {\bf 75}, 119908 (2007)].
  %%CITATION = PHRVA,D68,032002;%%

 \bibitem{otherexp}
   P.~Krokovny {\it et al.}  [Belle Collaboration],
  %``Observation of the D/sJ(2317) and D/sJ(2457) in B decays,''
  Phys.\ Rev.\ Lett.\  {\bf 91}, 262002 (2003);
%  [arXiv:hep-ex/0308019].
  %%CITATION = PRLTA,91,262002;%%
K.~Abe {\it et al.},
  %``Measurements of the D/sJ resonance properties,''
  Phys.\ Rev.\ Lett.\  {\bf 92}, 012002 (2004);
%  [arXiv:hep-ex/0307052].
  %%CITATION = PRLTA,92,012002;%%
 A.~Drutskoy {\it et al.}  [Belle Collaboration],
  %``Observation Of Anti-B0 $\to$ D*(Sj)(2317)+ K- Decay,''
  Phys.\ Rev.\ Lett.\  {\bf 94}, 061802 (2005);
  %%CITATION = PRLTA,94,061802;%%
 E.~W.~Vaandering  [FOCUS Collaboration],
  %``Charmed hadron spectroscopy from FOCUS,''
  arXiv:hep-ex/0406044;
  %%CITATION = HEP-EX/0406044;%%
 B.~Aubert {\it et al.}  [BABAR Collaboration],
  %``Study of $B \to D_{sJ}^{(*)+} \bar{D}^{(*)}$ decays,''
  Phys.\ Rev.\ Lett.\  {\bf 93}, 181801 (2004);
%  [arXiv:hep-ex/0408041].
  %%CITATION = PRLTA,93,181801;%%
  B.~Aubert {\it et al.}  [BABAR Collaboration],
  %``Observation of a narrow meson decaying to $D_s^+ \pi^0 \gamma$ at a mass of
  %2.458-GeV/c$^2$,''
  Phys.\ Rev.\  D {\bf 69}, 031101 (2004);
  %[arXiv:hep-ex/0310050].
  %%CITATION = PHRVA,D69,031101;%%
    B.~Aubert {\it et al.}  [BABAR Collaboration],
  %``Measurement of the D/sJ*(2317)+ and D/sJ(2460)+ properties in e+ e- -->  c
  %anti-c production,''
  arXiv:hep-ex/0408067.
  %%CITATION = HEP-EX/0408067;%%

\bibitem{Colangelo:2004vu}
For a review see: P.~Colangelo, F.~De Fazio and R.~Ferrandes,
  %``Excited charmed mesons: Observations, analyses and puzzles,''
  Mod.\ Phys.\ Lett.\  A {\bf 19}, 2083 (2004).
  %%CITATION = MPLAE,A19,2083;%%

\bibitem{fdf1}
 P.~Colangelo and F.~De Fazio,
  %``Understanding D/sJ(2317),''
  Phys.\ Lett.\  B {\bf 570}, 180 (2003);
%  [arXiv:hep-ph/0305140].
  %%CITATION = PHLTA,B570,180;%%
  P.~Colangelo, F.~De Fazio and A.~Ozpineci,
  %``Radiative transitions of D/sJ*(2317) and D/sJ(2460),''
  Phys.\ Rev.\  D {\bf 72}, 074004 (2005).
%  [arXiv:hep-ph/0505195].
  %%CITATION = PHRVA,D72,074004;%%

\bibitem{Aubert:2006mh}
  B.~Aubert  [BABAR Collaboration],
  %``Observation of a new D/s meson decaying to D K at a mass of
  %2.86-GeV/c**2,''
  Phys.\ Rev.\ Lett.\  {\bf 97}, 222001 (2006).
  %%CITATION = PRLTA,97,222001;%%

\bibitem{van Beveren:2006st}
  E.~van Beveren and G.~Rupp,
  %``New BABAR state D/sJ(2860) as the first radial excitation of the
  %D/s0*(2317),''
  Phys.\ Rev.\ Lett.\  {\bf 97}, 202001 (2006).
  %%CITATION = PRLTA,97,202001;%%

  \bibitem{swanson}
  F.~E.~Close, C.~E.~Thomas, O.~Lakhina and E.~S.~Swanson,
  %``Canonical Interpretation of the D/sJ(2860) and D/sJ(2690),''
  Phys.\ Lett.\  B {\bf 647}, 159 (2007).
  %%CITATION = PHLTA,B647,159;%%

  \bibitem{Colangelo:2006rq}
  P.~Colangelo, F.~De Fazio and S.~Nicotri,
  %``$D_{sJ}(2860)$ resonance and the $s_\ell^P={5\over 2}^-$ $c \bar s$ ($c
  %\bar q$) doublet,''
  Phys.\ Lett.\  B {\bf 642}, 48 (2006).
  %%CITATION = PHLTA,B642,48;%%

\bibitem{Collaboration:2007nr}
  J.~Koponen (UKQCD Collaboration),
  %``Energies of B_s meson excited states - a lattice study,''
  arXiv:0708.2807 [hep-lat].
  %%CITATION = ARXIV:0708.2807;%%

\bibitem{Neubert:1993mb}
 For a review see:  M.~Neubert,
  %``Heavy quark symmetry,''
  Phys.\ Rept.\  {\bf 245}, 259 (1994).
%  [arXiv:hep-ph/9306320].
  %%CITATION = PRPLC,245,259;%%

\bibitem{Abe:2003zm}
  K.~Abe {\it et al.}  [Belle Collaboration],
  %``Study of B- --> D**0 pi- (D**0 --> D(*)+ pi-) decays,''
  Phys.\ Rev.\  D {\bf 69}, 112002 (2004).
  %%CITATION = PHRVA,D69,112002;%%

  \bibitem{ferrandes}
  A  discussion can be found in:
   P.~Colangelo, F.~De Fazio and R.~Ferrandes,
  %``Bounding effective parameters in the chiral Lagrangian for excited  heavy
  %mesons,''
  Phys.\ Lett.\  B {\bf 634}, 235 (2006).
  %%CITATION = PHLTA,B634,235;%%

 \bibitem{traceformalism}
 %\cite{Georgi:1990um}
%\bibitem{Georgi:1990um}
  H.~Georgi,
  %``AN EFFECTIVE FIELD THEORY FOR HEAVY QUARKS AT LOW-ENERGIES,''
  Phys.\ Lett.\  B {\bf 240}, 447 (1990).
  %%CITATION = PHLTA,B240,447;%%
% Trace formalism (Georgi)

\bibitem{hqet_chir}
M.B.Wise, Phys. Rev. D {\bf  45}  R2188 (1992); G.Burdman and
J.F.Donoghue, Phys. Lett. B {\bf  280} 287 (1992); P.Cho, Phys.
Lett. B {\bf  285}  145 (1992); H.-Y.Cheng {\it et al.,}  Phys.
Rev.  D  {\bf 46}  1148 (1992); R.Casalbuoni {\it et al.,} Phys.
Lett.  B {\bf 299} 139 (1993).

\bibitem{Zhang:2006yj}
  B.~Zhang, X.~Liu, W.~Z.~Deng and S.~L.~Zhu,
  %``$D_{sJ}(2860)$ and $D_{sJ}(2715)$,''
  Eur.\ Phys.\ J.\  C {\bf 50}, 617 (2007).
  %[arXiv:hep-ph/0609013].
  %%CITATION = EPHJA,C50,617;%%
  
  \bibitem{Wei:2006wa}
  W.~Wei, X.~Liu and S.~L.~Zhu,
  %``D wave heavy mesons,''
  Phys.\ Rev.\  D {\bf 75}, 014013 (2007).
%  [arXiv:hep-ph/0612066].
  %%CITATION = PHRVA,D75,014013;%%
  
\bibitem{Colangelo:1995ph}
P.~Colangelo, G.~Nardulli, A.~Deandrea, N.~Di Bartolomeo, R.~Gatto
and F.~Feruglio,
  %``On the coupling of heavy mesons to pions in QCD,''
  Phys.\ Lett.\  B {\bf 339}, 151 (1994);
  %%CITATION = PHLTA,B339,151;%%
V.~M.~Belyaev, V.~M.~Braun, A.~Khodjamirian and R.~Ruckl,
  %``D* D pi and B* B pi couplings in QCD,''
  Phys.\ Rev.\  D {\bf 51}, 6177 (1995);
  %%CITATION = PHRVA,D51,6177;%%
  P.~Colangelo, F.~De Fazio, G.~Nardulli, N.~Di Bartolomeo and R.~Gatto,
  %``Strong coupling of excited heavy mesons,''
  Phys.\ Rev.\  D {\bf 52}, 6422 (1995);
  %%CITATION = PHRVA,D52,6422;%%
P.~Colangelo and F.~De Fazio,
  %``QCD interactions of heavy mesons with pions by light-cone sum rules,''
  Eur.\ Phys.\ J.\  C {\bf 4}, 503 (1998).
  %%CITATION = EPHJA,C4,503;%%


\bibitem{Neubert:1997uc}
 For a review  see: M.~Neubert and B.~Stech,
  %``Non-leptonic weak decays of B mesons,''
  Adv.\ Ser.\ Direct.\ High Energy Phys.\  {\bf 15}, 294 (1998).
  %%CITATION = 00319,15,294;%

\bibitem{PDG}
%\cite{Yao:2006px}
%\bibitem{Yao:2006px}
  W.~M.~Yao {\it et al.}  [Particle Data Group],
  %``Review of particle physics,''
  J.\ Phys.\ G {\bf 33}, 1 (2006).
  %%CITATION = JPHGB,G33,1;%%

\bibitem{slope1}
%\cite{Bowler:2002zh}
%\bibitem{Bowler:2002zh}
  K.~C.~Bowler, G.~Douglas, R.~D.~Kenway, G.~N.~Lacagnina and C.~M.~Maynard
                  [UKQCD Collaboration],
  %``Semi-leptonic decays of heavy mesons and the Isgur-Wise function in
  %quenched lattice QCD,''
  Nucl.\ Phys.\  B {\bf 637}, 293 (2002).
%  [arXiv:hep-lat/0202029].
  %%CITATION = NUPHA,B637,293;%%

  \bibitem{slope2}
  B.~Aubert {\it et al.}  [BABAR Collaboration],
  %``Determination of the Form Factors for the Decay B0 -> D*- l+ nu_l and of
  %the CKM Matrix Element |V_cb|,''
  arXiv:0705.4008 [hep-ex].
  %%CITATION = ARXIV:0705.4008;%%

%\cite{Abe:2001cs}
\bibitem{Abe:2001cs}
  K.~Abe {\it et al.}  [BELLE Collaboration],
  %``Determination of |V(cb)| using the semileptonic decay anti-B0 --> D*+  e-
  %anti-nu,''
  Phys.\ Lett.\  B {\bf 526}, 247 (2002).
%  [arXiv:hep-ex/0111060].
  %%CITATION = PHLTA,B526,247;%%


 \bibitem{hfag}
   E.~Barberio {\it et al.}  [Heavy Flavor Averaging Group (HFAG)
                  Collaboration],
  %``Averages of b-hadron properties at the end of 2006,''
  arXiv:0704.3575 [hep-ex],
  %%CITATION = ARXIV:0704.3575;%%
with online update at  http://www.slac.stanford.edu/xorg/hfag/


\bibitem{Pedlar:2007za}
  T.~K.~Pedlar {\it et al.}  [CLEO Collaboration],
  %``Measurement of B(D/s+ --> l+ nu) and the decay constant f(D/s+),''
  arXiv:0704.0437 [hep-ex].
  %%CITATION = ARXIV:0704.0437;%%
 \end{thebibliography}
\end{document}